\documentclass[10pt, conference, compsocconf]{IEEEtran}
  \usepackage{amssymb}
  \usepackage{latexsym}
  \usepackage{amsfonts}
  \usepackage{amsthm}
  \usepackage{amsmath}
  \usepackage{graphics}
  \usepackage{setspace}
  \usepackage{graphicx}
  \usepackage{epstopdf}
  \usepackage{color}
  \usepackage{float}
  \usepackage{wrapfig}
  \usepackage{color}
  \usepackage{rotating}
  \usepackage{array} 
  \usepackage{dblfloatfix}
  \usepackage[hyphens]{url}
  \usepackage[table]{xcolor}
  \usepackage{multirow}

  \usepackage{graphics}
  \usepackage{setspace}
  \usepackage{algorithm}
  \usepackage{algorithmic}
  \usepackage{graphicx}
  \usepackage{epstopdf}  
  \usepackage{color}
  \usepackage{float}
  \usepackage{wrapfig}
  \usepackage{color}

\newenvironment{noindlist}
 {\begin{list}{\labelitemi}{\leftmargin=1.2em \itemindent=-.5em}}
 {\end{list}}

\newcommand{\comment}[1]{}

\ifCLASSINFOpdf
\else
\fi
\begin{document}
%
\title{Context-aware Sensor Search, Selection and Ranking Model for Internet of Things Middleware}




%


\author{\IEEEauthorblockN{Charith Perera\IEEEauthorrefmark{1}\IEEEauthorrefmark{2},
 Arkady Zaslavsky\IEEEauthorrefmark{2},
Peter Christen\IEEEauthorrefmark{1},
Michael Compton\IEEEauthorrefmark{2} and
Dimitrios Georgakopoulos\IEEEauthorrefmark{2}}
\IEEEauthorblockA{\IEEEauthorrefmark{1}Research School of Computer Science, The Australian National University, Canberra, ACT 0200, Australia}
\IEEEauthorblockA{\IEEEauthorrefmark{2}CSIRO ICT Center, Canberra, ACT 2601, Australia}}

%


\maketitle

\begin{abstract}
As we are moving towards the Internet of Things (IoT), the number of sensors deployed around the world is growing at a rapid pace. Market research has shown a significant growth of sensor deployments over the past decade and has predicted a substantial acceleration of the growth rate in the future. It is also evident that the increasing number of IoT middleware solutions are developed in both research and commercial environments. However, sensor search and selection remain a critical requirement and a challenge. In this paper, we present \textit{CASSARAM}, a context-aware sensor search, selection, and ranking model for Internet of Things to address the research challenges of selecting sensors when large numbers of sensors with overlapping and sometimes redundant functionality are available. CASSARAM proposes the search and selection of sensors based on user priorities. \textit{CASSARAM} considers a broad range of characteristics of sensors for search such as reliability, accuracy, battery life just to name a few. Our approach utilises both semantic querying and quantitative reasoning techniques. User priority based weighted Euclidean distance comparison in multidimensional space technique is used to index and rank sensors. Our objectives are to highlight the importance of sensor search in IoT paradigm, identify important  characteristics of both sensors and data acquisition processes which help to select sensors, understand how semantic and statistical reasoning can be combined together to address this problem in an efficient manner. We developed a tool called \textit{CASSARA} to evaluate the proposed model in terms of resource consumption and response time.

\end{abstract}


%
\IEEEpeerreviewmaketitle

\section{Introduction}
\label{sec:Introduction}

The numbers of sensors deployed around the world are increasing at a rapid pace. These sensors continuously generate enormous amounts of data. Collecting and storing data from all the available sensors may not create additional value or solve the problem of efficient sensor data processing. Further, it may not be feasible due to large scale, resource limitations, and cost factors. When significant amounts of sensors are available to choose from, it becomes a challenge and a time consuming task to select the \textit{appropriate} sensors. We describe the term \textit{appropriate} in section \ref{sec:Problem_Definition_and_Motivation}.

Sensing as a service (SensaaS) model is expected to build on top of the IoT infrastructure and services. SensaaS model envisions that sensors and/or sensor data streams would be available to use over the Internet following some utility arrangements. Currently, several middleware solutions that are expected to facilitate such model are under development. OpenIoT \cite{P377}, GSN \cite{P227}, Cosm \cite{P579} are some examples. These middleware solutions strongly focus on connecting sensor devices to software system and related functionalities \cite{P377}. However, when more and more sensors get connected to the Internet, the sensor search functionality becomes critical.

This paper addresses the growing challenge of sensor search and selection in IoT solutions and research. Traditional web search approach will not work in IoT sensor selection and search domain as text based search approaches cannot capture the critical characteristics of a sensor accurately. Another approach that can be followed is meta data annotation (e.g. basic details related to each sensor such as sensor type, manufacturer, capability). Even if we maintain meta data on sensors (e.g. stored in sensor's storage) or in the cloud, interoperability will be a significant issue. Further, a user study done by Broring et al. \cite{P586} has described an approach where 20 participants were asked to enter metadata for a weather station sensor using a simple user interface. Those 20 persons made 45 mistakes in total. The requirement of re-entering metadata in different places (e.g. enter metadata into GSN once and again enter metadata into OpenIoT) arises when we do not have common descriptions.

Recently, W3C Incubator Group released the Semantic Sensor Network XG Final Report that defines SSNO ontology \cite{P581}. SSNO describes sensors, their characteristics and relationships between concepts. SSNO strengthens sensor interoperability and accuracy avoiding error-prone manual data entry. Further, inconsistencies in sensor descriptions can be avoided by letting the sensor hardware manufactures to produce and make available sensor descriptions using ontologies so that IoT solution developers can retrieve and incorporate (e.g. mapping) them into their system. Ontology based sensor description and data modelling is useful for IoT solutions. This approach also allows semantic querying. Our proposed solution allows the users to express their priorities in terms of sensor characteristics and it will search and select appropriate sensors. In our model, both quantitative reasoning  and semantic querying techniques are employed to improve the efficiency and performance of the system by utilizing strengths of both techniques.

The rest of this paper is structured as follows: In Section \ref{sec:Background}, we highlight our vision and where the findings of this paper is going to be fit in. We describe CA4IOT (Context Awareness for Internet of Things) architecture in brief in order to emphasize the importance of this research. At the end, we briefly review relevant literature. Next, we explain problem definitions and motivation in Section \ref{sec:Problem_Definition_and_Motivation}. At the end, we list main research contributions of this paper. In Section \ref{sec:Real_World_Challenge}, we discuss a real world application related to agricultural domain. Our proposed solution, CASSARAM, is presented in detail in Section \ref{sec:Our_Solution}. Data models, context frameworks, algorithms, and architectures are also discussed. In Section \ref{sec:Implementation_and_Experimentations}, we provide implementation details including tools, software platforms,  hardware platforms, and data sets used in this work. We also discuss  assumptions made during the experiments and results of the experiments. Discussion on research findings is presented in Section \ref{sec:Evaluations}. Finally, we present conclusions and prospects for future work in Section \ref{sec:Conclusions and Future Work}.


\begin{table*}[ht!]
\centering
\footnotesize
\caption{Comparison of web services selection and sensors selection Domains}
\begin{tabular}{ m{0.05cm} m{8cm} m{8cm}  }

\hline  
   \multicolumn{1}{r}{}   &    
\begin{center} Web Service Selection Domain \end{center} & 
\begin{center} Sensor Selection Domain \end{center} 

\\[-0.25cm] \hline \hline


\begin{sideways}Similarities\end{sideways}     
&   \begin{noindlist}
 		 \item Consuming single web service may not create significant value. Therefore, web services selection and composition is critical to generate value.
 		 \item Many alternative web services are available to use
 		 \item Can be found through directory services
 		 \item Quality of services matters
                 \item There are free as well as paid services

    \end{noindlist}  
 &  \begin{noindlist}
  		 \item Collecting data from a single sensor may not create significant value. Therefore, sensor selection and composition is critical to generate value.
                 \item Many alternative sensors will be available to use
                 \item Middleware solutions such as OpenIoT and GSN will play a mediator roles between sensors  and sensor data consumers 
                 \item Quality of sensors (and data) matters
                 \item There will be free as well as paid sensors
 		 
     \end{noindlist}  

\\ 
\begin{sideways} Differences  \end{sideways}      
&   \begin{noindlist}
		 \item Largely guided by standards
                 \item Largely depend on software
                 \item Less uncertainty (unless some hardware resources are involved. e.g. weather information.)
                 \item Not tangible and more reliable
                 \item Some web services accept data as input and produce some data based on them (e.g. data fusion)
                 \item Data send to the consumer using web services
                 \item Comparatively, less number of web services will be available to access over Internet by 2020
                 \item Typically provide more meaningful processed and refined data.

\end{noindlist}  
 &
  \begin{noindlist}
 		 \item No standards (yet)
                 \item Largely depend on hardware and firmware
                 \item More uncertainty
                 \item Tangible, could be mobile and less reliable
                 \item Some sensors may accept queries/conditions/preferences as inputs and produce data based on them. Nevertheless, sensors do not accept raw data with the intention of fusing data.
                 \item Data will send to the consumer using different techniques such as web services, IP packets, http protocols, etc.
		 \item Comparatively, large number of sensors will be available to access over Internet by 2020
                 \item Typically provides less meaningful raw sensor data
 \end{noindlist}  

\\[-0.25cm] \hline

\end{tabular}

\label{Tbl:Sensor Search and Webservices Domain}
\vspace{-0.33cm}
\end{table*}


\section{Background and Related Work}
\label{sec:Background}

In this section, we briefly explain our broader vision of automated (context-aware) configuration of filtering, fusion and reasoning mechanisms, according to the problems/tasks at hand in IoT paradigm. This will be a significant enhancement over current paradigms, where resources are configured statically and a priori. This vision is part of OpenIoT project \cite{P377}. In our previous work, we explained the complete architecture called CA4IOT \cite{ZMP004} that facilitates above vision. Here, we briefly introduce CA4IOT and explain how CASSARAM would fit and benefit from CA4IOT vision.


Manually selecting and configuring relevant sensors and data fusion operators when large numbers of sensors are available to use is not feasible \cite{ZMP004} or is very hard. For example, users such as environmental scientists (i.e. non-technical personnel) may not have sufficient knowledge in computer science but they need to retrieve sensor data for their work. They are only interested in acquiring relevant data so they can use the data to build models, simulations, understand and solve their problems.

\begin{figure}[h]
 \centering
 \includegraphics[scale=.53]{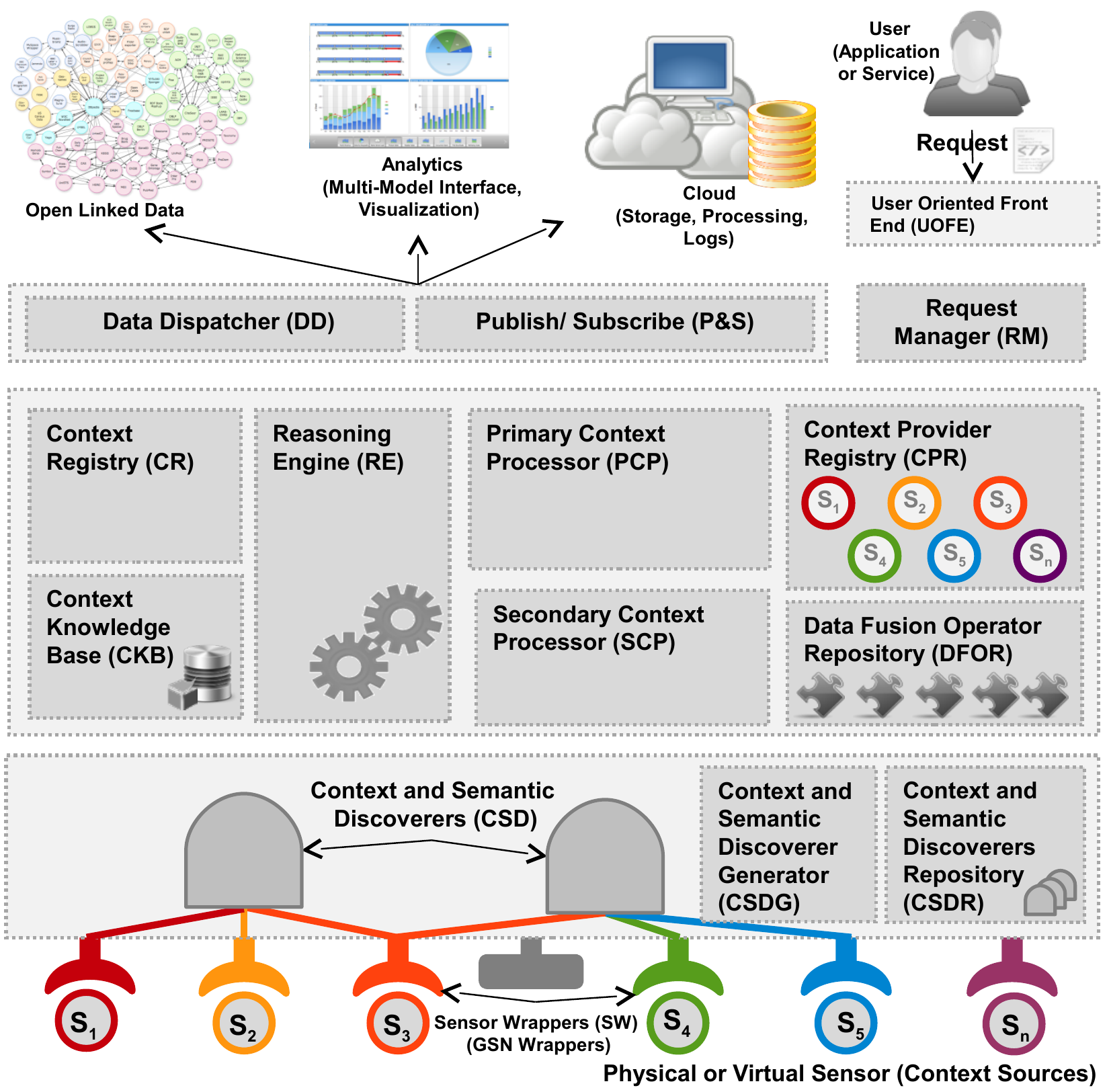}
\vspace{-0.63cm}	
 \caption{CA4IOT Architecture supports \textit{sensing as a service} model. Detailed description of this architecture and the execution process is presented in \cite{ZMP004}.}
 \label{Figure:Execution_and_interaction_process_of_CA4IOT}	
\vspace{-0.53cm}	
\end{figure}

Let us consider a scenario where an environmental scientist wants to measure environmental pollution in Canberra. There is no single sensor that is capable of measuring environmental pollution. For example, environmental pollution can be roughly attributed to three sub categories: land pollution, air pollution, and water pollution. Each category can be measured by a large number of (different types) sensors. Furthermore, there is high level context information that may not become available by processing data retrieved from a single sensor directly. Data retrieved from multiple sensors need to be fused together dynamically at run time to generate such high level context. Manual selection of sensors and data fusion operators in order to facilitate automated sensor configuration could be complex in IoT due to its scale and dynamic nature. Therefore \textit{``How to efficiently select appropriate sensors  by understanding the user requirements /problems despite inherent complexity of the challenge?''} is the problem we addressed in CA4IOT. CA4IOT helps the users by automating the task of selecting sensors according to problems/tasks at hand. It focuses on  breaking down the user requirements and understanding which sensors can provide relevant information to the users. Once this is completed, CA4IOT needs to find the most suitable sensors and that is exactly where the thrust of this paper fits in. This problem becomes more challenging when alternatives are available. (e.g. User wants to gather data from 100 temperature sensors located in specific area where 25000 sensors are available while each sensor has different characteristics in terms of accuracy, reliability, cost, etc.). CA4IOT vision is illustrated in  Figure \ref{Figure:Execution_and_interaction_process_of_CA4IOT}.


\begin{figure}[h]
 \centering
 \includegraphics[scale=.75]{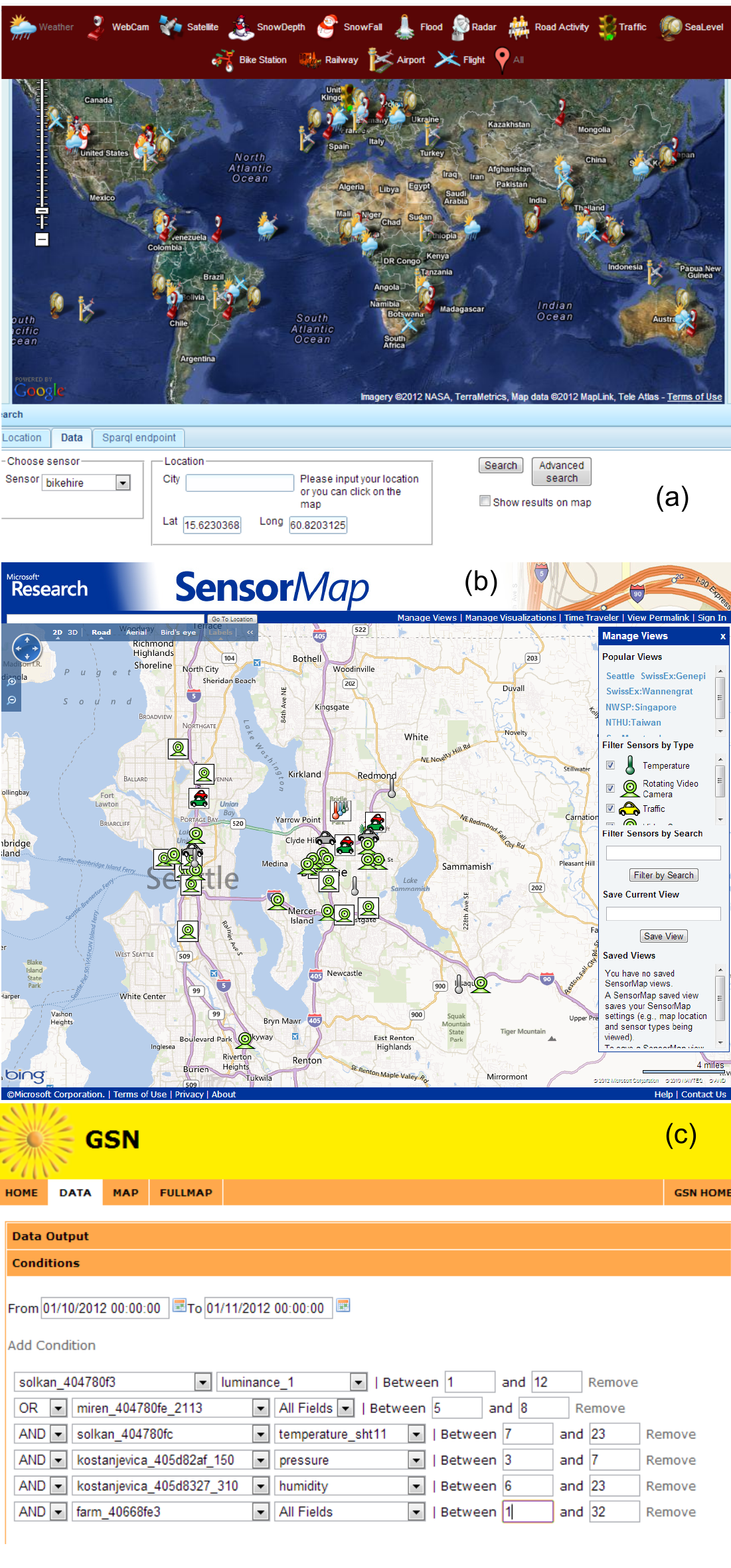}
\vspace{-0.50cm}	
 \caption{Some of the leading IoT middelware solutions. Only limited searching functionality is provided  across all solutions. (a) Linked Sensor Middleware (LSM) by Digital Enterprise Research Institute \cite{P583,P584}, (b) SensorMap by Microsoft \cite{P578}, Global Sensor Network (GSN) initiated by  EPFL \cite{P227}.}
 
 \label{Figure:Samples1}	
\vspace{-0.50cm}	
\end{figure}

 Ideally, IoT middleware solutions should empower users to express what they want and provide the relevant sensor data back to the users quickly without asking the users to manually select sensors which are relevant to their requirements. Even though IoT has received significant attention both in academia and industry, sensor search and selection have not been addressed properly. The following examples show how existing IoT middleware solutions provide sensor searching functionality.

Linked Sensor Middleware (LSM) \cite{P583,P584} provides some sensor selection and searching functionality. However, LSM has limited capabilities such as selecting sensors based on location and sensor types. All the searching needs to be done using SPARQL query language which is not very intuitive. Similar to LSM, there are several other IoT middleware related projects under development. GSN \cite{P227}  is a platform aimed at providing flexible middleware to address the challenges of sensor data integration and distributed query
processing. It is a generic data stream processing engine. GSN has gone beyond the traditional sensor network functionality such as routing, data aggregation, and energy optimisation. GSN lists all the available sensors in a combo-box which users need to select. Another approach is Microsoft SensorMap \cite{P578}. It only allows users to select sensors by using a location map, by sensor type and by keywords. COSM (formerly Pachube) \cite{P579} is another approach which provides a secure, scalable platform that connects devices and products with applications to provide real-time control and data storage. COSM also offers only keyword search. Our proposed solution CASSARAM can be used to complement the above mentioned IoT middleware solutions with extensive sensor search and selection functionality. Figure \ref{Figure:Samples1} shows some of the leading IoT middleware solutions.


 In Table \ref{Tbl:Sensor Search and Webservices Domain}, we present a comparison of similarities and differences between sensor selection and web services selection domains. According to a study in Europe [2], there are over 12,000 working and useful Web services on the Web. Even in such conditions, choice between alternatives (depending on context properties) has become a challenging problem. The similarities strengthen the argument that sensor selection is an important challenge at the same level of complexity as web services. On the other hand, differences show that sensor selection will become a much more complex challenge  over the coming decade due to the scale of IoT.

 In the following, we briefly describe some of the work done in sensor searching and selection. Truong et al. \cite{P587} propose a fuzzy based similarity score comparison sensor search technique to compare output of a given sensor with outputs of several other sensors to find out a matching sensor. Mayer et al. \cite{P588} considers location of smart things/sensors as the main context property and structures them in a logical structure. Then, sensors are searched by location using tree search techniques. Search queries are distributively processed in different paths/nodes of the tree. Elahi et al. \cite{P589} propose a content-based sensor search approach (i.e. finding a sensor that outputs a given value at the time of a query. Dyser is a search engine proposed by Ostermaier et al. \cite{P590} for real-time Internet of Things, which uses statistical models to make predictions about the state of its registered objects (sensors). When a user submits a query, Dyser pulls latest data to identify the actual current state to decide whether it matches the user query. Prediction models help to find matching sensors with minimum number of sensors data retrievals. Very few related efforts focus on sensor search based on context information. 


\section{Problem Definition and Motivation}
\label{sec:Problem_Definition_and_Motivation}

The problem that we address in this paper can be defined as following. Due to the increasing numbers of available sensors, we need to search and select sensors that provide data which will help to solve our problems at hand in the most efficient and effective way. In order to accomplish this task, we need to search and select sensors  based on context. Mainly, we identify two categories of requirements: point-based requirements (non-negotiable or mandatory) and proximity-based (negotiable or flexible) requirements. 

First, there are point-based requirements that need to be definitely fulfilled. For example, if a user is interested in measuring temperature in a certain location (e.g. Canberra), the result (e.g. list of sensors) should only contain sensors that can measure temperature. User cannot be satisfied by providing any other types of sensors (e.g. pressure sensors). There is no  alternative or compromise in this type of requirements. Location can be identified as a point-based requirement. The second category is proximity-based requirements where users may be willing to suffice with some variations or compromise. For example,  user has the same interest as before. However, in this situation, a user may impose proximity-based requirements in addition to the point-based requirements. User may like to have accuracy of the sensors around 92\%, and reliability 85\%. Therefore, the user gives highest priority  to these characteristics. User may accept sensors that closely fulfil his/her requirements even though all other characteristics may not be favourable (e.g. cost of acquisition may be high and sensor response time may be slow). It is important to note that users will not be able to provide any specific value so the system should be able to understand the user priorities and provide the results accordingly by using comparison techniques. One of the motivating arguments behind this research work are current market statistics and predictions that show rapid growth in sensor deployments related to IoT and Smart Cities. By 2020, there will  be 50 to 100 billion devices connected to the Internet \cite{P029}. Further, our work is motivated by increasing trend of IoT middleware solutions development. Today, most of the leading midddleware solutions provide only limited sensor search and selection functionality as depicted in Figure \ref{Figure:Samples1}.

\newpage
In this paper, we propose a model that can complement any IoT middleware solution. Our contributions can be summarised as follows.

\begin{itemize}
\item We propose a context based framework for sensors in IoT middleware which allows to capture and model sensor characteristics. This information allows users to search sensors based on context information.

\item We modelled our proposed context framework as an extension to Semantic Sensor Ontology (SSNO) and it is compatible with many existing SSNO-based developments. It is envisioned that the extended SSNO can be used in projects such as Phenonet \cite{P412} and OpenIoT \cite{P377}.

\item We propose CASSARAM that allows users to search and select sensors based on user priorities. Users are able to get not only the sensors that will provide required sensor data  but also sensors that have characteristics (i.e. context information related to sensors and sensor data acquisition process) that users prefer most. Our approach can be used in Sensing-as-a-Service approach.

\item We develop CASSARA tool that allows the users to express their priorities in a comparative manner. Slider UI components allow the comparison. CASSARA also populates a ranked list of sensors in order from best choice to worst.

\item We propose a novel technique called \textit{Comparative Priority-based Heuristic Filtering} algorithm to make the sensor indexing algorithms  faster and more efficient.

\item We evaluate CASSARAM using our prototype tool and measure performance and efficiency in terms of computational resource consumption.

\end{itemize}

\section{Real World Challenge}
\label{sec:Real_World_Challenge}

\begin{figure}[t]
 \centering
 \includegraphics[scale=.14]{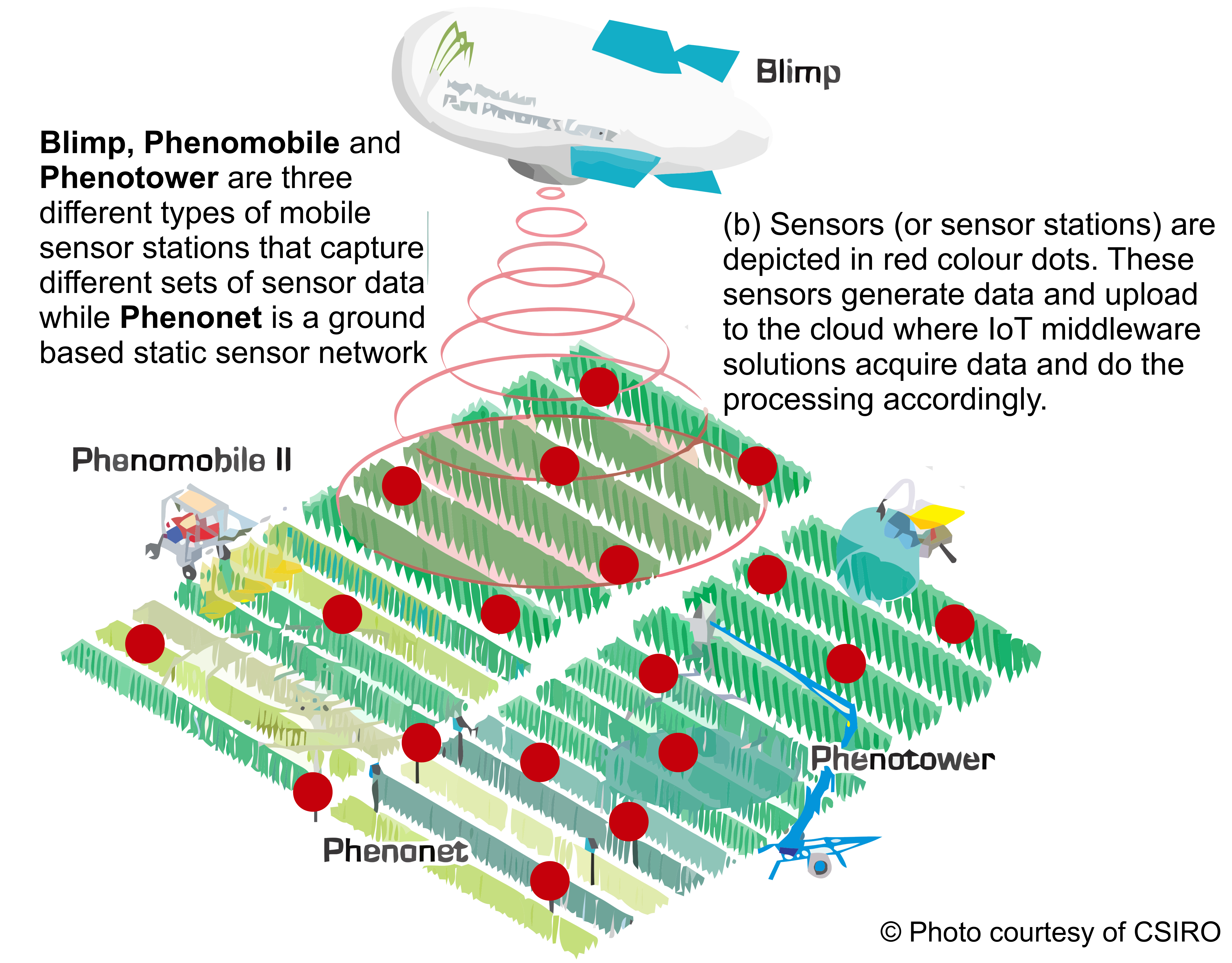}
\vspace{-0.33cm}	
 \caption{State of the art sensor based monitoring in agriculture domain.}
 \label{Figure:Real_World_Challenge}	
\vspace{-0.43cm}	
\end{figure}

In this section we present a real world example application to reinforce the arguments and to strengthen the necessity of addressing sensor search and selection  challenges. It will also help to understand the challenges more clearly. Figure \ref{Figure:Real_World_Challenge} shows state of the art sensor based monitoring system used in Australian agricultural domain. Australia is the fourth largest wheat and barley exporter after US, Canada and EU. There are two challenges that Australian agriculture has to address: scarcity of water resources and low soil fertility. Every year, Australian grain breeders plant up to 1 million 10m$^{2}$ plots across the country to find the best high yielding varieties of wheat and barley. The plots are usually located in remote places often requiring more than four hours travel one-way to reach. The challenge is to monitor the crop performance and growing environment through different seasons and return the information in an easily accessible format. The challenge of crop growing and performance monitoring can be addressed by deploying sensors. Querying the collected sensor data is essential to understand what is happening in the field. The challenge is to develop a sensor-searching model which allows to search sensors based on context information. As we mentioned earlier, it is not required to collect data from all the sensors deployed in all the plots all the time which is inefficient. For example, find out what sensors have more energy and collect data only from those sensors helps to run the entire network for much longer time without reconfiguring.

\section{CASSARAM: The Proposed Approach: }
\label{sec:Our_Solution}

In this section, we propose a sensor selection approach step-by-step. First, we provide a high-level overview of the model which describes the overall execution flow and critical steps of the model. Then, we explain how user preferences are captured. Next, data representation model and proposed extensions are presented. Semantic querying and quantitative reasoning techniques are discussed.

\begin{figure}[t!]
 \centering
 \includegraphics[scale=.81]{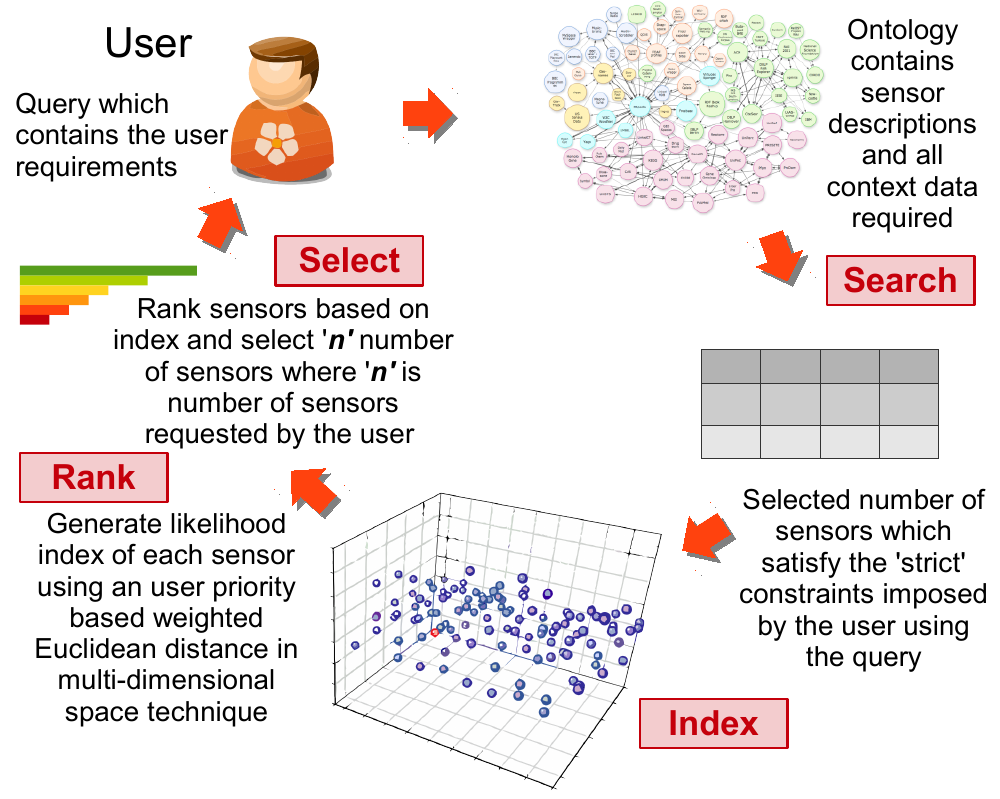}
\vspace{-0.33cm}	
 \caption{High level Overview of CASSARAM}
 \label{Figure:System_Architecture}	
\vspace{-0.60cm}	
\end{figure}

The critical steps of CASSARAM are presented in Figure \ref{Figure:System_Architecture}. As we mentioned earlier our objective is to allow the users to search and select sensors that best suit their requirements. In our model, we divide user requirements into two categories: \textit{point-based requirements} and \textit{proximity-based requirements}. Point-based requirements are also called non-negotiable requirements and they must be satisfied exactly as specified by the user. In contrast, proximity-based requirements, which are also called negotiable, may or may not be satisfied exactly as specified by the user. Results that are closest to the user requirements would be selected.

Algorithm \ref{Alg:System_Flow_of_CASSARA_Model} describes the execution flow of CASSARAM. Common algorithmic notations used in this paper are presented below:  \textit{Ontology} ($\mathbb{O}$) consists of sensor descriptions and context property values related to all sensors, \textit{UserPrioritySet} ($\mathbb{P}$) contains user priority values for all  context properties, \textit{Query} ($\mathbb{Q}$) consists of point-based requirements expressed in SPARQL, Number of sensors ($\mathbb{N}$) required by the user, \textit{ResultsSet} ($\mathbb{S}_{Results}$) contains selected number of sensors, IndexedSensorSet ($\mathbb{S}_{Indexed}$), Multidimensional Space ($\mathbb{M}$) where each context property is represented by a dimension, \textit{UserInput} ($\mathbb{UI}$) consists of input values provided to CASSARAM by the users via user interface, \textit{ScalingInformation} ($\mathbb{SC}$) defines the scale of the slider,  \textit{WeightedUserPrioritySet} ($\mathbb{P}^{Weighted}$) provides details on how user has prioritised context properties, \textit{ContextPropertySet} ($\mathbb{CP}$) consists of all context information, \textit{NormalizedContextPropertySet}  $(\mathbb{NCP})$.

 \begin{algorithm}[b]

 \algsetup{linenosize=}
 \caption{Execution Flow of CASSARAM}
 \label{Alg:System_Flow_of_CASSARA_Model}
 \begin{algorithmic}[1]
 \REQUIRE ($\mathbb{O}$), ($\mathbb{P}$), ($\mathbb{Q}$),   ($\mathbb{N}$), ($\mathbb{S}_{Results}$), ($\mathbb{S}_{Indexed}$), ($\mathbb{M}$).
  
 \STATE \textbf{Output:}  $\mathbb{S}_{Results}$
 \STATE $\mathbb{S}_{Filtered} \leftarrow queryOntology(\mathbb{O},\mathbb{Q})$
 \IF{$cardinality(\mathbb{S}_{Filtered})< \mathbb{N}$ }
 \RETURN $\mathbb{S}_{Results}  \leftarrow    \mathbb{S}_{Filtered} $
 \ELSE
 \STATE $\mathbb{P}  \leftarrow captureUserPriorities(\mathbb{UI})$ 
 \STATE $\mathbb{M}  \leftarrow $Plot Sensors in Multidimensional Space ($\mathbb{S}_{Results}$)
 \STATE $\mathbb{S}_{Indexed} \leftarrow calculateCPWI ({\mathbb{S}_{Results},\mathbb{M}})$
 \STATE $\mathbb{S}_{Results}  \leftarrow rankSensors(\mathbb{S}_{Indexed})$
  \STATE $\mathbb{S}_{Results} \leftarrow selectSensors(\mathbb{S}_{Results},\mathbb{N})$
 \RETURN $\mathbb{S}_{Results}$
 \ENDIF 
 \end{algorithmic}
 \end{algorithm}

At the beginning, CASSARAM identifies point-based requirements, proximity-based requirements, and user priorities. First, users need to select the point-based requirements. For example, a user may want to collect sensor data from 1000 temperature sensors deployed in Canberra. In this situation, sensor type (i.e.temperature), location (i.e. Canberra) and number of sensors required (i.e. 1000) are point-based requirements. Our CASSARA prototype tool  provides user interface to express this information via SPARQL queries. SPARQL query is dynamically built when users select the context properties. In CASSARAM, any context property can become a point-based requirement. Then users can define proximity-based requirements. All context properties we discuss in this section are available to be defined in comparative fashion by setting priorities via a slider-based user interface as depicted in Figure \ref{Figure:CASSARA_Tool_GUI}. Next, each sensor is plotted into a multidimensional space where each dimension is represented by a context property (e.g. accuracy, reliability, latency). Each dimension is normalized 0 to 1. Then, Comparative Priority-based Weighted Index (CPWI) is generated for each sensor combining user priorities and context properties. Finally, sensors are ranked according to CPWI and  the number of sensors required by the user are selected from the top of the list.

\begin{figure}[h]
 \centering
 \includegraphics[scale=.86]{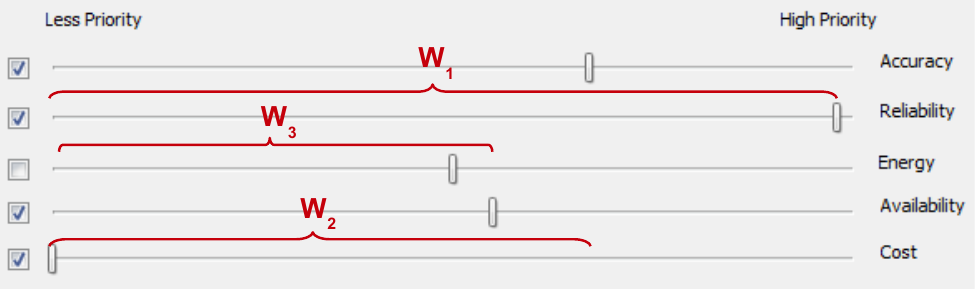}
\vspace{-0.13cm}	
 \caption{Weight $W_{1}$ is assigned to \textit{reliability} property. Weight $W_{2}$ is assigned to \textit{Accuracy} property. Weight $W_{3}$ is assigned to \textit{availability} property and finally weight $W_{4}$, default weight, is assigned to \textit{cost} property. High priority means always favourable and low priority means always unfavourable. For example, if user makes \textit{cost} a high priority (more towards right), that means CASSARAM tries to find the sensors that produce data at the lowest cost. Similarly, if user makes \textit{accuracy} a high priority, that means CASSARAM tries to find the sensors that produce data with high accuracy.}
 \label{Figure:User_Priority_Capturing}	
\vspace{-0.12cm}	
\end{figure}

User priority capturing (UPC)  is a technique we developed to capture user priorities through a user interface shown in Figure \ref{Figure:User_Priority_Capturing}. CASSARAM allows users to express what context property is more important to them compared to others. If a user does not want a specific context property to be considered in the indexing process, they can avoid it by not selecting the check-box correlating to that specific context property. For example, according to Figure \ref{Figure:User_Priority_Capturing}, \textit{energy} will not be considered when calculating CPWI. This means user is willing to accept sensors with any \textit{energy} consumption level.

%

As depicted in Figure \ref{Figure:User_Priority_Capturing}, if users want more reliable sensors to be ranked higher compared to accuracy of the sensors, the reliability slider need to be placed more to the right compared to the accuracy slider. A weight is calculated for each context property. Therefore, more priority means higher weight. As a result, sensors with high reliability and accuracy will be ranked higher. However, those sensors may have high costs due to low priority placed on cost property.


In this paper, we use Semantic Sensor Network Ontology (SSNO) \cite{P581} to model sensor descriptions and context properties. Main reasons to select SSNO are interoperability and the trend moving towards ontology usage in IoT and sensor data management domain. A comparison of different semantic sensors ontologies are presented in \cite{P103}. The SSNO is capable of modelling significant amounts of information about sensors such as sensor capabilities, performance, the conditions in which sensors can be used, etc. Details are presented in \cite{P581}. SSNO includes  most common context properties such as accuracy, precision, drift, sensitivity, selectivity, measurement range, detection limit, response time, frequency and latency. SSNO can be extended unlimitedly by  sub classing three classes: \textit{measurement property}, \textit{operating property}, and \textit{survival property}. We extend the \textit{quality class} by adding several sub-classes based in order to facilitate our context framework.

In the next step, which we call "Ranking Using Quantitative Reasoning" sensors are ranked based on proximity-based user requirements. We developed a weighted Euclidean distance-based indexing techniques called Comparative Priority-based Weighted Index (CPWI)  as follows.

\begin{center}
$\left ( CPWI \right ) = \sqrt{\sum_{i=1}^{n} \left [  W_{i}(U_{i}^{d} -  S_{i}^{\alpha })^{2}\right ]    } $
\end{center}

\begin{figure}[h]
 \centering
 \includegraphics[scale=.52]{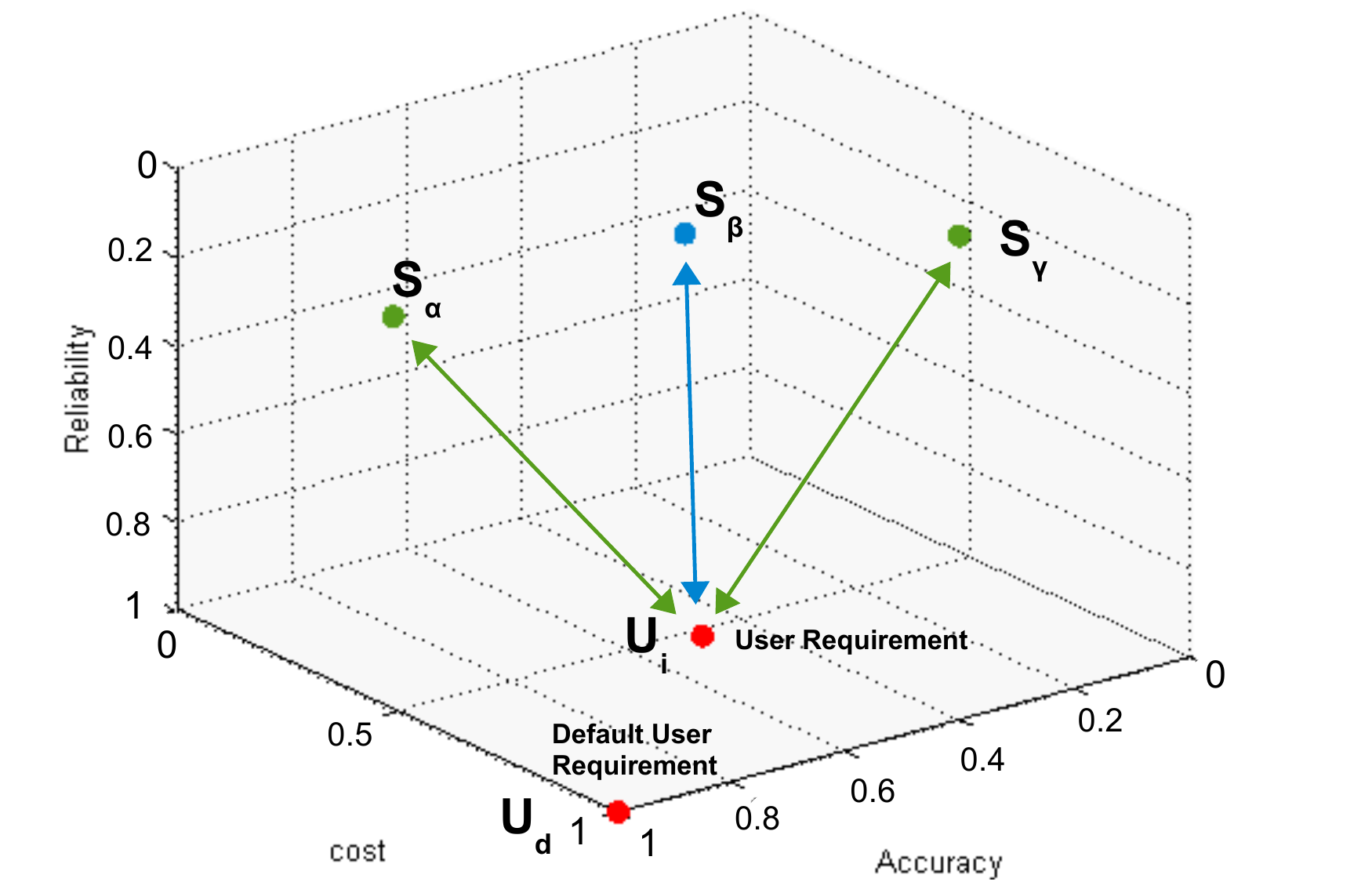}
 \vspace{-0.63cm}	
 \caption{Sensors plotted in three-dimensional space for demonstration purposes. $S_{\alpha}$, $S_{\beta}$, and $S_{\gamma}$ represent real sensors. CPWI calculate weighted distance between $S_{j}$ and $U_{i||d}$. Shortest distance means sensor will rank higher because it is close to the user requirement. }
 \label{Figure:Sensor_Plot}	
 \vspace{-0.15cm}	
 \end{figure}

First, each sensor is plotted in multidimensional space where each context property is represented by a dimension. Then, users can plot an ideal sensor in the multidimensional space by manually entering context property values as illustrated in Figure \ref{Figure:Sensor_Plot} by $U_{i}$. By default, CASSARAM will automatically plot an ideal sensor as depicted in $U_{d}$ (i.e. highest value for all context properties).  Next, user priorities are retrieved and weights are calculated in comparative fashion.


 
%
%
 
We use the following context framework (i.e. context properties listed below) after evaluating a number of research proposals carried out in the areas of quality of service domain related web services \cite{P593}, mobile computing \cite{P592}, and sensor ontologies \cite{P581}, we extracted following context properties to be stored and maintained about each sensor. This information helps to decide which sensor to be used in a given situation. We adopt the following definition to our work. \textit{``Context is any information that can be used to characterise the situation of an entity. An entity is a person, place, or object that is considered relevant to the interaction between a user and an application, including the user and applications themselves.''}\cite{P104}. CASSARAM has no constraints on a number of context properties that can be used.  Our context framework comprises availability, accuracy, reliability, response time, frequency, sensitivity, measurement range, selectivity, precision, latency, drift, resolution, detection limit, operating power range, system (sensor) lifetime, battery life, security, accessibility, robustness, exception handling, interoperability, configurability, user satisfaction rating, capacity, throughput, cost of data transmission, cost of data generation, data ownership cost, bandwidth, and trust. The list is extendible with more context attributes if necessary.

The solution we discussed so far works well with small number of sensors. However, model becomes inefficient when the number of sensors available to search increases. Let us consider an example to identify the inefficiency. Assume we have access to one million sensors. A user wants to select 1,000 sensors out of them. In such situation, CASSARAM will index and rank one million sensors using proximity-based requirements provided by the user and select top 1,000 sensors. However, indexing and ranking all possible sensors (in this case one million) is inefficient and wastes significant amount of computational resources. Further, CASSARAM will not be able to process large number of user queries due to such inefficiency. We propose a technique called Comparative Priority-based Heuristic Filtering (CPHF) to make CASSARAM more efficient.  The basic idea is to remove sensors that are positioned far away from user defined ideal sensor and reduce the number of sensors that need to be indexed and ranked.

Consider the above scenario. First, all the eligible sensors rank in descending order of the highest weighted context property (in this case accuracy). Then, remove 40\% (from $\mathbb{N}_{Removable}$) of the sensors from the bottom of the list. Then order the remaining sensors in descending order of the next highest weighted context property (in this case reliability).  Then, remove 30\% (from $\mathbb{N}_{Removable}$) of the sensors from the bottom of the list. This process applies for the remaining context properties as well. Finally, index and rank the remain sensors. This approach dramatically reduces the indexing and ranking related inefficiencies. Broadly, this category of techniques are called Top-K selection where top sensors are selected in each iteration.

\section{Implementation and Experimentations}
\label{sec:Implementation_and_Experimentations}

The aim of implementation and experimentation is to study the performance of CASSARAM in different IoT related scenarios which we developed based on real world requirements. Experimentation setup, datasets used, assumptions, experiment testbed, and results are presented.

\begin{figure}[h]
\centering
\includegraphics[scale=.62]{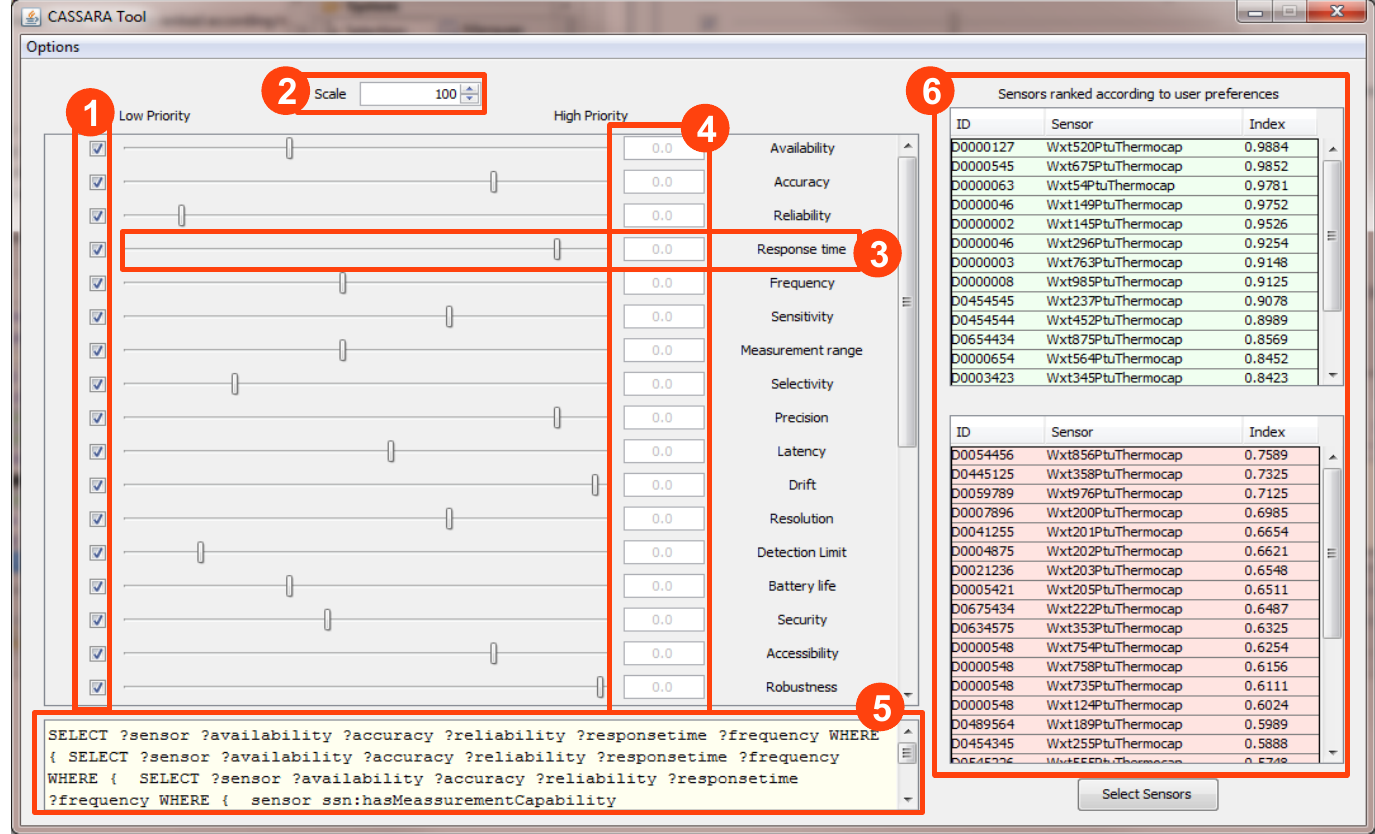}
\vspace{-0.23cm}	
\caption{User interface of the CASSARA tool. (1) Users need to check the box to express that they do concern about that specific context property, (2) Allows to set the scale. Slider becomes more sensitive when scale increased, (3) Slider attach to each context property can be configured to express the priority in comparative fashion, (4) Ideal value related to each context property can be entered. Defaults is zero, (5) Allows to enter SPARQL query that consists point-based requirement. However, by default, tool generates the SPARQL appropriately based on the context properties selected by the users, (6) Shows list of sensors ranked according to index.}
\label{Figure:CASSARA_Tool_GUI}	
\vspace{-0.23cm}	
\end{figure}

The proposed model analysed and evaluated using a prototype which we developed using Java is called \textit{`CASSARA Tool'}. The data was stored in MySQL database. As shown in Figure \ref{Figure:CASSARA_Tool_GUI}, it allows to capture user preferences regarding their expected priorities on each characteristic of a sensor. We used a computer with Intel(R) Core i5-2557M 1.70GHz CPU  and 4GB RAM to evaluate our proposed model. In order to perform mathematical operations such as Euclidean distance calculation in multidimensional space, we used  Apache Commons mathematics \cite{P580} library. It is an open source optimized library of lightweight, self-contained mathematics and statistics components addressing the most common problems not available in the Java programming language. As we used Semantic Sensor Ontology (SSNO) \cite{P581} to manage sensor descriptions and related data, we employed open source Apache Jena API \cite{P436} to process and manipulate semantic data. We conducted each experiment 100 times and averages are taken into account.

Our evaluation used a combination of real data and synthetically generated data. We collected environmental linked-data from the bureau of meteorology \cite{P582} and data sets from  both Phenonet project \cite{P412}  and  Linked Sensor Middleware (LSM) project \cite{P583,P584}. The main reasons to combine data are due to the need of generating a large amount of data  and the need of controlling different aspects of data (e.g. context information related to sensors need to be embedded into the data set, because real data that matches our context framework is not available in any public data sets at the moment) to better understand the behaviour of CASSARAM in different IoT related real world situations and scenarios where real data is not available.

We make the following assumptions in our work. We assume that sensor descriptions such as sensor capabilities and measurements  are already retrieved from sensor manufacturers and merged into SSNO. Similarly, we assume that context data related to each sensor such as current power level, power consumption, accuracy, reliability are retrieved by software systems that manage such data and are available to be used. We acknowledge these data could be stored in a distributed manner (e.g. each GSN instance may contain descriptions of sensors which are connected to that specific instance). Therefore, sensor search may need to be performed in a distributed manner. However, we do not consider distributed aspect in this work and leave it for future work.

In order to evaluate CASSARAM, we used a data set which comprises sensor descriptions  and context information  for \textit{one million sensors}. We synthetically created this data set by combining different real data sets. In this section, we present experimentation results with  brief explanations on each graph. Interpretations of each graph and overall discussions are provided in Section \ref{sec:Evaluations}.

\begin{figure}[h]
 \centering
 \vspace{-0.33cm}
 \includegraphics[scale=.53]{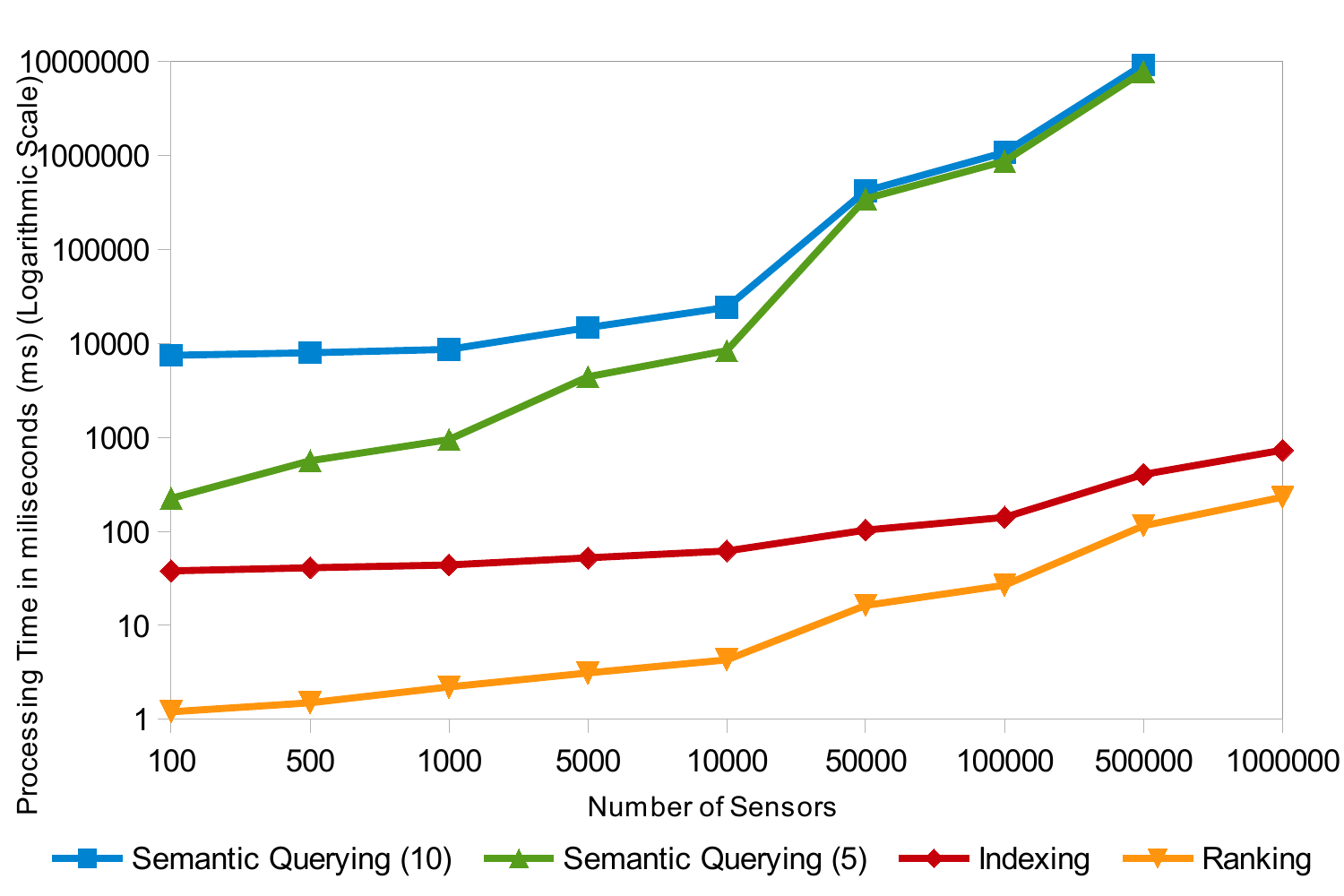}
\vspace{-0.33cm}	
 \caption{This graph shows processing time taken by each step and the total  sensor  selection process when the number of sensors gets increased. The number of context properties used for indexing kept at 30 for ranking and indexing experiences. Semantic querying use 5 and 10 context properties. Note: Y-axis is measured in milliseconds and presented in logarithmic scale.}
 \label{Figure:Results1}	
\vspace{0.13cm}	
\end{figure}

\begin{figure}[h]
 \centering
 \vspace{-0.33cm}
 \includegraphics[scale=.53]{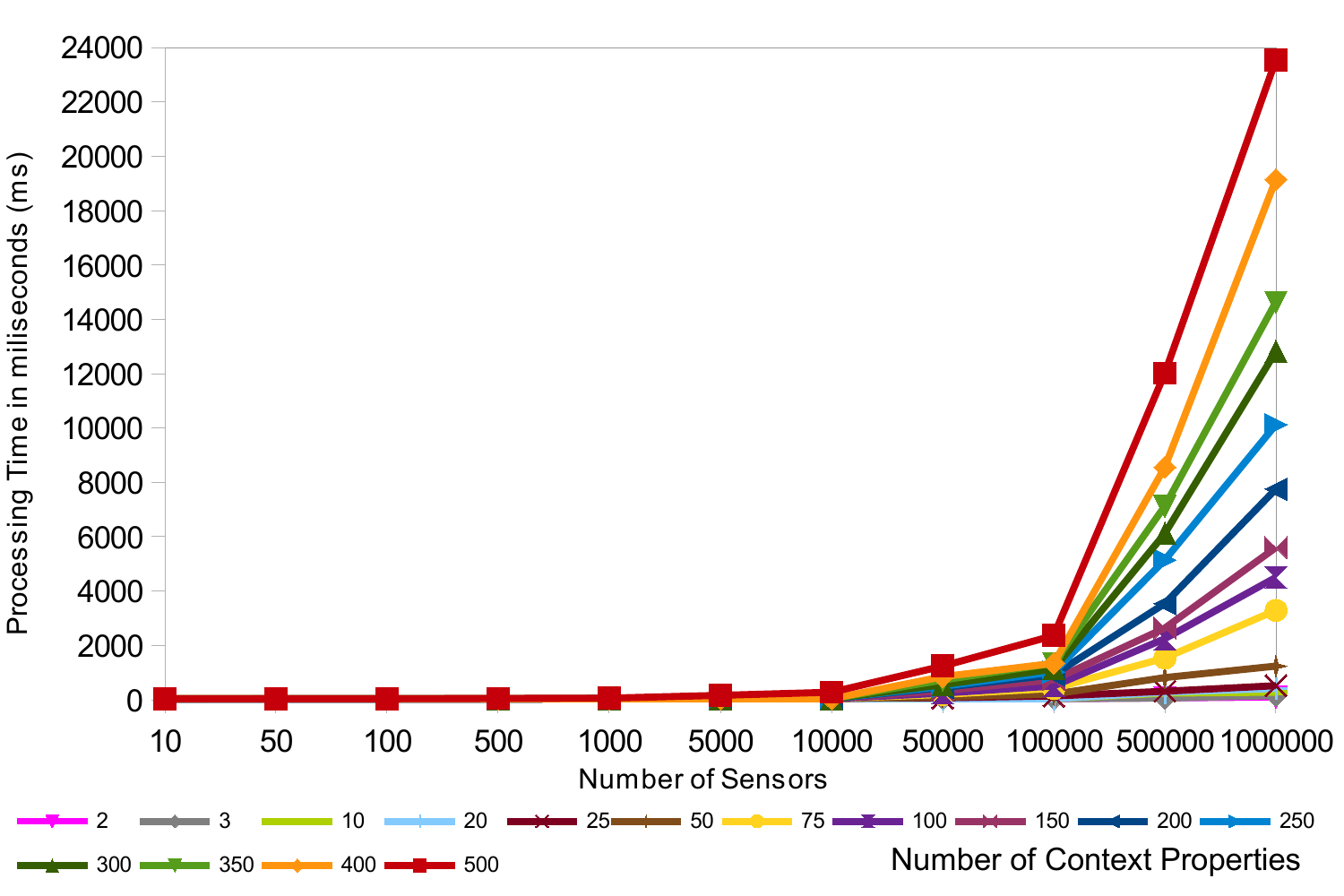}
\vspace{-0.33cm}	
 \caption{This graph shows processing time taken by sensor indexing process when number of context properties and number of sensors get increased. Synthetic context properties are used to for experiments.}
 \label{Figure:Results2}	
\vspace{-0.33cm}	
\end{figure}

\begin{figure}[h]
 \centering
	
 \includegraphics[scale=.53]{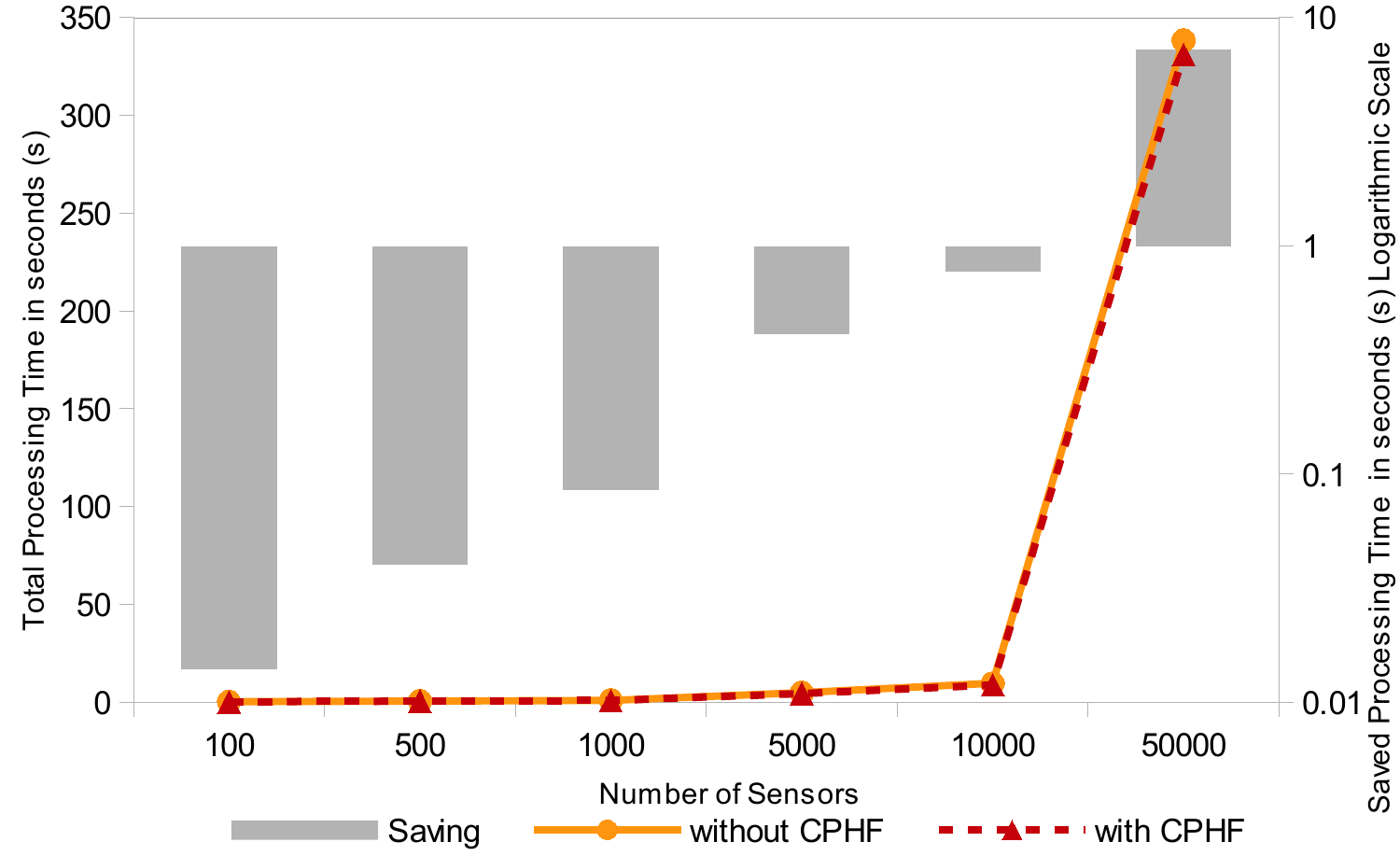}
\vspace{-0.33cm}	
 \caption{This graph compares time taken by sensor selection process with and without CPHF algorithm when number of sensors get increased. Number of sensors that users require kept at 50 in all experiments ($\mathbb{N}$=50). 30 context properties are retrieved via semantic querying, indexed and ranked.}
 \label{Figure:Results3}	
\vspace{0.13cm}	
\end{figure}

\begin{figure}[h!]
 \centering
 \includegraphics[scale=.53]{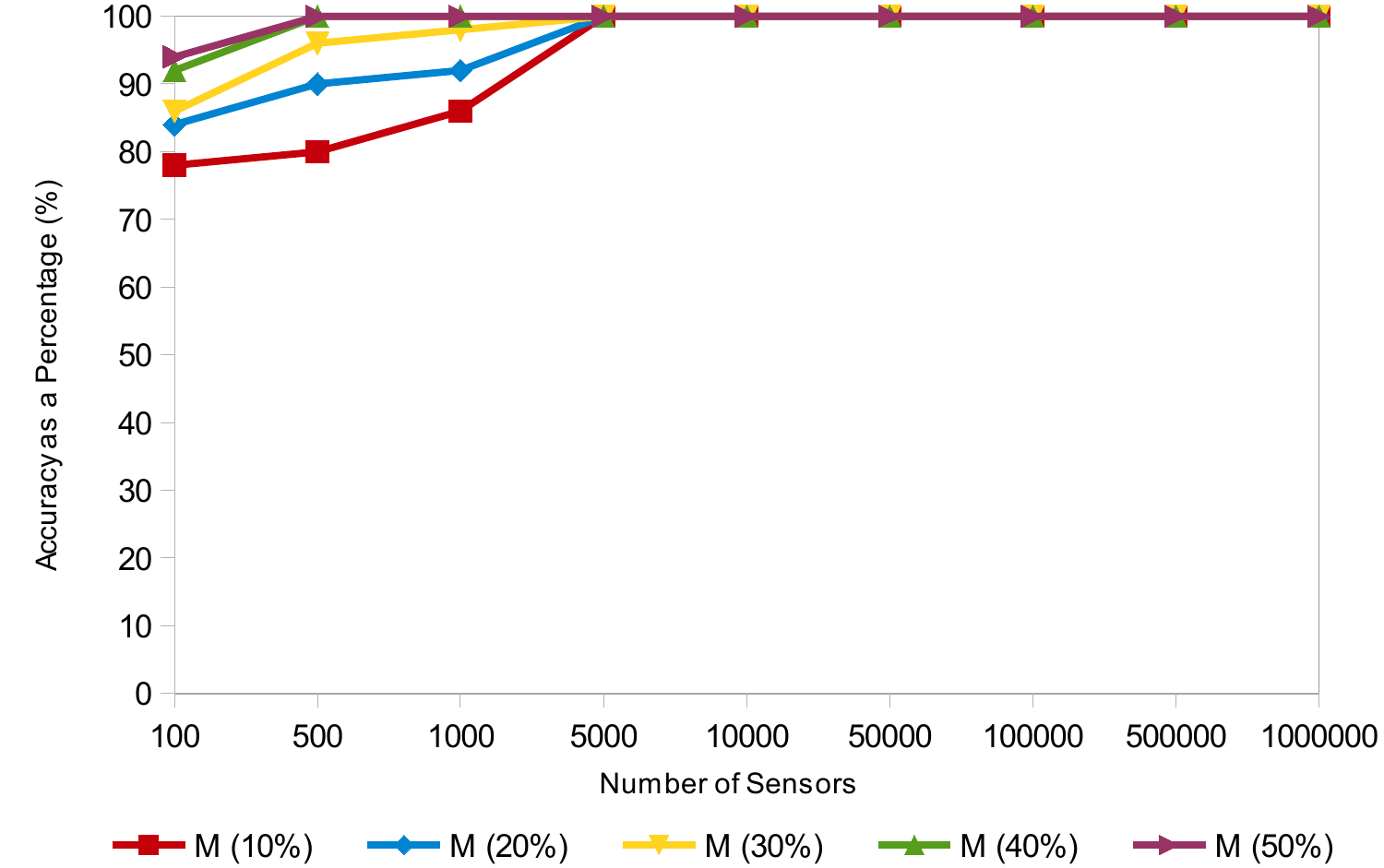}
\vspace{-0.33cm}	
 \caption{This graph shows how accuracy changes when Margin of Error (M\%) value changes in CPHF algorithm and number of sensors get increased. Number of sensors that users require kept at 50 in all experiments ($\mathbb{N}$=50.) }
 \label{Figure:Results4}	
\vspace{-0.63cm}	
\end{figure}


\section{Evaluation and Discussion}
\label{sec:Evaluations}

As depicted in Figure \ref{Figure:Results1}, semantic querying consumes significantly more processing time compared to indexing and  ranking. Further, when the number of context properties that are retrieved by a query increases, execution time also increases significantly. MySQL can  join only 61 tables which only allows to retrieve maximum of 10 context properties in SSNO data structure. Alternative data storage or running multiple queries (can be efficient when having fewer sensors. see comparison between semantic querying times ) can be used as an alternative solution. We conclude that optimising semantic querying will make significant impact on overall performance of CASSARAM. As depicted in Figure \ref{Figure:Results2}, reducing the number of indexed sensors below 10,000 allows to perform CASSARAM faster. Processing time starts to get increased significantly after 100,000 sensors. As depicted in Figure \ref{Figure:Results3}, the complexity of CPHF (due to sub queries) has not effected the total processing time significantly. Instead, CPHF saved some amount of time in indexing and ranking phases. CPHF method returns only limited number of sensors where non-CPHF approach returns all the sensors available to CASSARAM which consumes more resources  including more processing time, significant amount of memory and temporary storage. According to Figure \ref{Figure:Results4}, accuracy of CPHF approach increases when margin of error (\textit{M}) increases. However, lower \textit{M} leads CASSRAM towards low resource consumption. Therefore, it is a trade-off between accuracy and resource consumption. The optimum value of \textit{M} can be dynamically learned by machine learning techniques based on what context properties are prioritized by the users in each situation and how the  normalized weights are distributed across different context properties. We will investigate these possibilities in future research.

\section{Conclusions and Future Work}
\label{sec:Conclusions and Future Work}

With the advances in sensor hardware technology and cheap materials, sensors are expected to be embedded into many objects around us which increases the number of sensors available to us. This means that we have access to multiple sensors that would measure similar environmental phenomenon. We need to  decide what operational and conceptual sensor related context properties are more important than others.

In this work, we showed how context information related to each sensor can be used to search and select sensors that are best suited for user requirements. We  selected sensors based on  user expectations and priorities. As proof of concept, we built a working prototype to demonstrate the functionality of our CASSARAM and to support experimentations in realistic applications. We also highlight how CASSARAM will help us to achieve our broader Sensing-as-a-Service vision in IoT paradigm. In future, we plan to incorporate CASSARAM  into leading IoT middleware solutions such as GSN, SensMA, and OpenIoT  to support automated sensor selection functionality in distributed environment. This will help us to perform more evaluations and understand how automated sensor selection would complement IoT middleware solutions. Further, we will investigate how semantic and quantitative reasoning can work together more closely to achieve efficient results and to provide more functionality.

\section{ACKNOWLEDGEMENT}
\label{sec:ACKNOWLEDGEMENT}

Authors acknowledge support from SSN TCP, CSIRO, Australia and ICT OpenIoT Project (Open source blueprint for large scale selforganizing
cloud environments for IoT applications), which is co-funded
by the European Commission under seventh framework
program, contract number FP7-ICT-2011-7-287305-OpenIoT. The
Author(s) acknowledge help and contributions from all partners of
the OpenIoT project.

\def\IEEEbibitemsep{0pt plus .5pt}
\bibliography{Bibliography}
\bibliographystyle{IEEEtran}

\end{document}